\documentclass[letter]{ieice}

\usepackage[T1]{fontenc}
\usepackage{txfonts}
\usepackage[implicit=false]{hyperref}

\newcommand{\ket}[1]{|#1\rangle}

\field{A}
\vol{100}
\no{2}
\begin{document}
\title{Quantum  Optimal Multiple Assignment Scheme for
Realizing General Access Structure of Secret Sharing}
 \authorlist{%
 \authorentry{Ryutaroh MATSUMOTO}{S}{labelA}
}
\affiliate[labelA]{The author is with the Department of
Information  and Communications Engineering, Tokyo Institute of Technology, 152-8550 Japan}
\received{2016}{7}{1}
\revised{2016}{9}{23}
\maketitle
 \begin{summary}
The multiple assignment scheme is to assign one or more shares
to single participant so that any kind of access structure
can be realized by classical secret sharing schemes.
We propose its quantum version including
\emph{ramp} secret sharing schemes.
Then we propose an integer optimization approach to
minimize the average share size.
\end{summary}

\section{Introduction}
Secret sharing (SS) \cite{shamir79} is a cryptographic scheme to
encode a secret to multiple shares being distributed to
participants, so that only \emph{qualified} sets of participants
can reconstruct the original secret from their shares.
Traditionally both secret and shares were classical information
(bits). Several authors, e.g.\ \cite{cleve99,gottesman00,smith00}
extended the traditional SS to quantum one
so that a quantum secret can be encoded to quantum shares.

A set of participants is called \emph{forbidden}
if the set has absolutely no information about the secret.
A secret sharing scheme is called \emph{perfect} \cite{stinson06}
 if
every set of participants is always qualified or forbidden.
If a set is neither qualified or forbidden in a secret sharing scheme,
the scheme is said to be \emph{ramp} or \emph{non-perfect}.
A merit of the ramp schemes is to reduce share size
(the number of bits or qubits) 
while keeping the secret size \cite{yamamoto86,ogawa05,ogata93}.

Traditionally, the access structure called the threshold
structure has been the most focused one, e.g.\ 
\cite{shamir79,cleve99,ogawa05},
where a set of participants is qualified if and only if
the number of participants is $\geq t$.
A scheme with a threshold structure is called a threshold scheme.
A well-known method to realize an arbitrary access structure is
the multiple assignment scheme proposed by Shamir \cite{shamir79}
and named by Ito et al.\ \cite{ito93}.
On the other hand, Smith \cite{smith00} showed how to
realize an arbitrary access structure in quantum \emph{perfect}
secret sharing schemes, while nobody has shown a construction
of quantum \emph{ramp} schemes with arbitrary access structures.

The multiple assignment scheme
assigns multiple shares of a threshold scheme
to single participants, and a single share can be
assigned to multiple participants.
It is not straight forward to adapt the multiple assignment
scheme, as the no-cloning theorem \cite{chuangnielsen}
prevents us from making multiple copies of a single quantum share.
The first purpose of this paper is to propose a quantum version
of the multiple assignment
scheme.

For a given size of secret,
it is important to reduce the size of shares.
A demerit of multiple assignment scheme in \cite{ito93}
was lack of consideration of share size.
Later, Iwamoto et al.\ \cite{iwamoto07} proposed
an integer optimization approach to minimize the worst-case or
the average share size of multiple assignment scheme.
The second purpose of this paper is to adapt Iwamoto et al.'s
integer optimization problem to our proposed quantum setting.

\section{Review of Previous Research Results}
By a classical secret sharing scheme,
we mean that its secret and its shares are classical information,
while by a quantum secret sharing scheme,
its secret and its shares are quantum information.
For a set $T$, $2^T$ denotes its power set
$\{ T_0 \mid T_0$ is a subset of $T\}$, and we have $|2^T| = 2^{|T|}$.

Firstly, we review the multiple assignment scheme
named by Ito et al.\ \cite{ito93} and originally
proposed by Shamir \cite{shamir79}.
The multiple assignment scheme construct
a classical secret sharing scheme with
$n$ participants from that with $m$ participants.
It is a map $\Phi$ from $\{1$, \ldots, $n\}$
to $2^{\{1, \ldots, m\}}$.
Let $W_1$, \ldots, $W_m$ be the shares of the original
secret sharing scheme.
The new secret sharing scheme constructed by $\Phi$
distributes $\{ W_j \mid j \in \Phi(i) \}$
to the $i$-th participants.
For example, suppose that $m=3$, $n=2$,
$\Phi(1)=\{1,2\}$, and $\Phi(2) = \{2,3\}$.
Then, in the new constructed secret sharing scheme,
the first participant receives $\{W_1$, $W_2\}$ as his/her share,
and the second one receives $\{W_2$, $W_3\}$ as his/her share.
This method works fine with the classical information.
But its straightforward extension to the quantum information
is \emph{impossible}, because the quantum no-cloning theorem
\cite{chuangnielsen} prevents us from distributing the
same $W_2$ to both first and second participants.
To avoid this impossibility,
we will focus the relation between
two shares $\{W_1$, $W_2\}$ and $\{W_2$, $W_3\}$,
which can be expressed by linear codes,
and will propose to transfer the relation to the quantum setting.

A classical secret sharing is said to be linear if
any linear combination of shares expresses the corresponding
linear combination of secrets \cite{chen07}.
Let $\mathbf{F}_q$ be a finite field with $q$ elements.
It was shown that any linear classical secret sharing scheme can be
expressed \cite[Proposition 1]{umberto15} by a pair of linear codes
$C_2 \subset C_1 \subset \mathbf{F}_q^n$ as follows,
provided that the linearity is considered over $\mathbf{F}_q$.

For a classical secret sharing scheme corresponding to $C_2 \subset C_1$,
the set of secrets is the factor space
\[
C_1/C_2 = \{ \vec{a} + C_2 \mid \vec{a}\in C_1\}.
\]
Therefore a secret $S  \in C_1 / C_2$ is a subset of $C_1$.
For a given secret $S \in C_1 / C_2$,
a vector $X = (X_1$, \ldots, $X_n)$
is chosen uniformly randomly from $S$.
A subset of $\{X_1$, \ldots, $X_n\}$ is distributed to each
participant as his/her share.
For practical use of secret sharing schemes,
it is indispensable to have a criterion by which
one can identify qualified or forbidden sets of shares.
Let $I \subset \{1$, \ldots, $n\}$,
and a set of participants collectively
have $\{ X_i \mid i \in I\}$ as their shares.
Let $P_I$ be the projection map sending
$(x_1$, \ldots, $x_n) \in \mathbf{F}_q^n$
to $(x_i)_{i\in I}$,
and $P_I(C_1) = \{ P_I(\vec{x}) \mid \vec{x} \in C_1 \}$.
It was shown in \cite{geil14} that
the set of shares expressed by $I$
is qualified iff
\begin{equation}
  \dim P_I(C_1)-\dim P_I(C_2) =
  \dim C_1 - \dim C_2, \label{eq:cq}
\end{equation}
and is forbidden iff
\begin{equation}
  \dim P_I(C_1)-\dim P_I(C_2) = 0. \label{eq:cf}
\end{equation}

It is known that most of quantum ramp secret sharing schemes
can also be described by 
a pair of linear codes $C_2 \subset C_1 \subset \mathbf{F}_q^n$ as follows
\cite{matsumoto14qss}.
Let $L = \dim C_1 - \dim C_2$,
then the dimension of (pure state) quantum secret is $q^L$
and its orthonormal basis can be chosen as
$\{ \ket{\vec{s}} \mid \vec{s} \in \mathbf{F}_q^L \}$.
The linear space of all the possible quantum shares
is $q^n$-dimensional, and its orthonormal basis can be chosen
as $\{ \ket{\vec{x}} \mid \vec{x} \in \mathbf{F}_q^n \}$.
We fix an $\mathbf{F}_q$-linear map $f$ from
$\mathbf{F}_q^L$ to the factor linear space $C_1/C_2$, and
a quantum secret $\ket{\vec{s}}$ is encoded to
\begin{equation}
\frac{1}{\sqrt{q^L}} \sum_{\vec{x} \in f(\vec{s})} \ket{\vec{x}},
\label{eq:qshare}
\end{equation}
which is the same as the encoding procedure of the CSS
quantum error-correcting codes \cite{calderbank96,steane96}.
Equation {(\ref{eq:qshare})} can be regarded as
a quantum state of $n$ particles having dimension $q$.
In this paper \emph{qudit} refers to a quantum object
that is represented by $q$-dimensional complex linear space.
Each participant receives a non-overlapping subset of the
$n$ particles of Eq.\ {(\ref{eq:qshare})} as his/her quantum
share.

As well as the classical case,
we need a criterion to tell if a set of shares is qualified
or forbidden. Recall that a share set is qualified if and
only if its complement is forbidden \cite{cleve99,ogawa05}
when the quantum secret sharing scheme is a pure-state scheme,
which encode a pure-state secret to a pure-state shares \cite{cleve99}.
It was also shown \cite{cleve99} that it is sufficient to consider
pure-state schemes.
Let $I \subset \{1$, \ldots, $n\}$,
and a set of participants collectively
have $I$ as their shares,
that is, the set of participants has the $i$-th quantum particle among
$n$ particles,
each of which has dimension $q$, if and only if $i \in I$.
The share set $I$ is qualified if and only if $I$ is qualified
and $\overline{I}$ is forbidden in the classical secret sharing
scheme constructed from $C_1 \supset C_2$, where $\overline{I}
=\{1$, \ldots, $n\}$.
In other words, $I$ is qualified if and only if both
Eq.\ (\ref{eq:cf}) with $I$ substituted by $\overline{I}$ and
Eq.\ (\ref{eq:cq}) hold.

\section{Proposed Method to Construct a Quantum Ramp Secret Sharing
  Scheme with a General Access Structure}
Suppose that the number of participants is $n$.
Let $q$ be a prime power as before, and
the dimension of quantum secret is assumed to be $q^L$
for a positive integer $L$.
Let $\mathcal{A}_Q \subset 2^{\{1, \ldots, n\}}$
be the family of qualified sets given as the requirement for
a quantum secret sharing scheme to be constructed.
Since we have restricted ourselves to the pure-state schemes,
the family of forbidden sets must be $\mathcal{A}_F =
2^{\{1, \ldots, n\}} \setminus \mathcal{A}_Q$.
It is also assumed that $\mathcal{A}_Q$ satisfies the
monotonicity condition \cite{stinson06},
that is,
if $A \in \mathcal{A}_Q$ and $A \subseteq B \in 2^{\{1, \ldots, n\}}$
then $B \in \mathcal{A}_Q$.
The monotonicity condition of $\mathcal{A}_Q$
implies the monotonicity condition of $\mathcal{A}_F$ in the reverse
order, that is, 
if $B \in \mathcal{A}_F$ and $B \supseteq  A \in 2^{\{1, \ldots, n\}}$
then $A \in \mathcal{A}_F$.

We introduce some notations from \cite{iwamoto07}.
Let $\vec{y} = (t$, $x_1$, \ldots, $x_{2^n-1})$.
Later $t$ becomes the design parameter of the underlying threshold
ramp quantum secret sharing scheme.
{Specifically, the underlying ramp secret sharing 
allows reconstruction of the secret only from $t$ or more shares.}
Let $b(p)_i$ as the $i$-th bit of the binary representation of
a positive integer $p$.
For a set $A \subset \{1$, \ldots, $n\}$,
define
\[
1(A)_p = \left\{
\begin{array}{ll}
  1&\textrm{if there exists }i\in A\textrm{ with }b(p)_i = 1,\\
  0&\textrm{otherwise}.
\end{array}
\right.
\]
Define $2^n$-dimensional vector
$a(\ell, A) = (\ell, 1(A)_1$, \ldots, $1(A)_{2^n-1})$.
Let $h_p$ be the number of $1$'s in the binary representation of
a positive integer $p$, and
$\vec{h}=(h_0$, $h_1$, \ldots, $h_{2^n-1})$.
As $\textrm{IP}^{R2}_{\widetilde{\rho}}$ in \cite{iwamoto07},
we solve the following integer optimization problem:
\[
\begin{array}{lc}
  \textrm{minimize}& \langle \vec{h}, \vec{y}\rangle,\\
  \textrm{subject to}& \langle a(-1, A), \vec{y}\rangle \geq 0,
  \forall A \in \mathrm{A}_Q,\\
 & \langle -a(-1, A), \vec{y}\rangle \geq L,
  \forall A \in \mathrm{A}_F,\\
  & \vec{y} \geq 0,
\end{array}
\]
where $\langle \cdot , \cdot \rangle$ denotes the inner product of
two vectors.
Since the above integer optimization problem is a relaxed version of
the original $\textrm{IP}^{R2}_{\widetilde{\rho}}$ in \cite{iwamoto07},
by Theorem 25 of \cite{iwamoto07}
there must be at least one solution $\vec{y}$ to our integer optimization
problem.
By following \cite{iwamoto07},
one can construct a \emph{classical} ramp secret sharing scheme
with $n$ participants,
the qualified set $\mathcal{A}_Q$, the forbidden set $\mathcal{A}_F$
and the classical secret consisting of $L$ symbols in $\mathbf{F}_q$.
Since their construction produces a classical linear secret sharing
scheme, it can be described by a nested pair of linear codes
$C_2 \subset C_1 \subset \mathbf{F}_q^m$,
where $m = x_1 + \cdots + x_{2^n-1}$ determined by
{a solution} $\vec{y} = (t$, $x_1$, \ldots, $x_{2^n-1})$
{of the above integer optimization problem.
In the construction method \cite{iwamoto07},
$(t, L, m)$  classical ramp secret sharing scheme is the 
underlying secret sharing scheme used in construction of the
desired secret sharing scheme.}
Observe that the constructed \emph{classical} ramp secret sharing
scheme has the minimum average share size, as proved in \cite{iwamoto07}.
Recall that in {the constructed} classical secret sharing scheme 
{expressed as}
$C_2 \subset C_1$, a participant receives some components of
$(z_1$, \ldots, $z_m) \in C_1$ as his/her share.
Let $V_i = \{ j \mid $ the $i$-th share contains $z_j \}$
{$\subset \{1$, \ldots, $m\}$.}

{To avoid violation of the quantum no-cloning theorem,
we modify the coding theoretic expression $C_1/C_2$ and
$V_i$. Note that only expression is modified and the secret sharing
scheme itself is not modified.
Let $\gamma(j) = |\{ i \mid j \in V_i \}|$, and
$m' = \gamma(1) + \cdots + \gamma(n)$.
For $\vec{z} = (z_1$, \ldots, $z_m) \in C_1$,
define 
\[
\phi(\vec{z}) = (\underbrace{z_1, \ldots, z_1}_{\gamma(1)\textrm{ times}}, \ldots, 
\underbrace{z_m, \ldots, z_m}_{\gamma(m)\textrm{ times}})\in \mathbf{F}_q^{m'}.
\]
Let $C'_1 = \phi(C_1)$ and 
$C'_2 = \phi(C_2)$.
Define $V'_i \subset \{1$, \ldots, $m'\}$ such that
$\{ z_\ell \mid \ell \in V_i \} =
\{ \phi(\vec{z})_j \mid j \in V'_i \}$
and $V'_i \cap V'_{i'} = \emptyset$ for $i \neq i'$,
where $\vec{z} = (z_1$, \ldots, $z_m) \in C_1$.
By the disjointedness $V'_i \cap V'_{i'} = \emptyset$
we can avoid the violation of the quantum no-cloning theorem.
The above change of notations makes no change in the actual
operation of the constructed classical secret sharing scheme.
Then the same constructed secret sharing scheme can also be
described by the code pair $C'_2 \subset C'_1$.
We also have $|V'_1| + |V'_2| + \cdots + |V'_n| = m'$.}

We construct the ramp \emph{quantum} secret sharing scheme from
{$C'_2 \subset C'_1 \subset \mathbf{F}_q^{m'}$}
in which the $i$-th participants receives the $j$-th qudit
among {$m'$} qudits if and only if {$w_j \in V'_i$,
where $(w_1$, \ldots, $w_{m'}) \in C'_1$ and
$m'$ qudits are defined from the quantum secret as Eq.\ (\ref{eq:qshare}).}
Then the constructed quantum secret sharing scheme has
$\mathcal{A}_Q$ as its qualified set,
$\mathcal{A}_F$ as its forbidden set,
and $L$-qudit in its quantum share,
and the average share size is generally small.

\section*{Acknowledgment}
The author would like to thank
Prof.\ Mitsugu Iwamoto for helpful discussion,
and an anonymous
reviewer to identify the critical error in the initial
manuscript.
This research is partly supported by the National
Institute of Information and Communications Technology,
Japan, and by the Japan Society for the Promotion of Science Grant
Nos.\ 23246071 and 26289116.


\end{document}